# In situ observation of slow and tunnelling light at the cutoff wavelength of an optical fiber

## YONG YANG AND M. SUMETSKY

*Aston Institute of Photonic Technologies, Aston University, Birmingham B4 7ET, UK*
*y.yang28@aston.ac.uk, m.sumetsky@aston.ac.uk*

**Slow waves and tunneling waves can meet at the cutoff wavelengths and/or transmission band edges of optical and quantum mechanical waveguides. The experimental investigation of this phenomenon, previously performed using various optical microstructures, is challenged by fabrication imperfections and material losses. Here, we demonstrate this phenomenon in situ for whispering gallery modes slowly propagating along a standard optical fiber, which possesses the record uniformity and exceptionally small transmission losses. Slow axial propagation dramatically increases the longitudinal wavelength of light and allows us to measure nanosecond-long tunneling times along tunable potential barriers having the width of hundreds of microns. This demonstration paves a simple and versatile way to investigate and employ the interplaying slow and tunneling light.**

In one-dimensional quantum mechanics [1], slow waves and tunneling waves meet at the top of a rectangular potential barrier as illustrated in Fig. 1(a). Energies smaller than the barrier height, $E < V_0$, correspond to tunneling, while energies $E > V_0$ correspond to the free propagation. At $E = V_0$, the propagation constant $\beta(E) \sim (E - V_0)^{1/2}$ is zero and, thus, tunneling and slowly propagating waves meet. For briefness, we call this phenomenon the STPC (Slow to Tunneling Propagation Crossover).

Similarly, the STPC can be observed for an electromagnetic wave which propagates along the direction $z$ of a uniform waveguide and varies as $\exp(i\beta(\lambda)z)$, where, in the vicinity of a cutoff wavelength (or transmission band edge) $\lambda_c$, the propagation constant $\beta(\lambda)$ depends on wavelength $\lambda$ as

$$\beta(\lambda) = \beta_0 + K\left(\lambda_c - \lambda + i\gamma\right)^{1/2}. \quad (1)$$

Here $\gamma$ determines the waveguide losses and $\beta_0$, $K$ are constants (see e.g., [2-5]). For example, for the whispering gallery mode (WGM) optical fiber waveguide considered in this Letter, we have $\beta_0 = 0$ and $K = 2^{3/2}\pi n_r \lambda_c^{-3/2}$ where $n_r$ is the refractive index of the fiber [5]. Due to the presence of losses, the tunneling and slow light propagation are not completely separated in the spectrum but overlap in the bandwidth $|\lambda - \lambda_c| < \Delta\lambda_{STPC}$. From Eq. (1), we have $\Delta\lambda_{STPC} > \gamma$. Experimental study of STPC for uniform waveguides with the dispersion relation approximated by Eq. (1) is of great interest since it allows us to simultaneously investigate the propagation of tunneling light [6-9], and slow light [10-13] in the most simple and straightforward way.

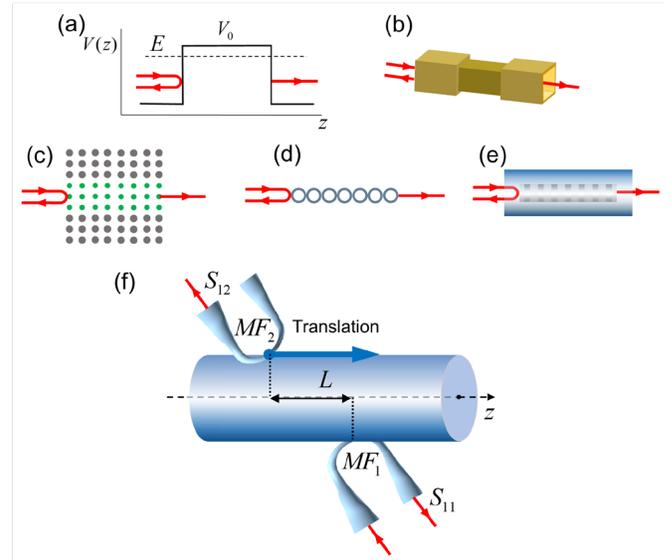

Fig. 1. (a) Potential barrier in quantum mechanics. (b) Microwave waveguide. (c) Photonic crystal waveguide. (d) Coupled microresonator waveguide. (e) Fiber Bragg grating waveguide. (f) Optical fiber WGM waveguide with two microfiber tapers coupled to this waveguide considered in the present paper.

The optical and microwave waveguides which can exhibit the STPC effect are illustrated in Fig. 1. The microwave waveguide shown in Fig. 1(b) contains a narrowed region, which corresponds to an effective potential barrier [2]. Figs. 1(c) (d), and (e) show optical waveguides formed by the periodic variation of the refractive index: a photonic crystal waveguide [10], a coupled microresonator waveguide [11, 14, 15], and a fiber Bragg grating (FBG) waveguide [16]. Finally, Fig. 1(f) shows the whispering gallery mode (WGM) waveguide based on a conventional optical fiber, which is considered in the present Letter.

Several challenging conditions are required for the clear observation of the STPC phenomenon. First, the propagation loss of the waveguide $\gamma$ should be as small as possible. This will allow us to achieve longer tunneling and free propagation times. However, small waveguide loss $\gamma$, or equivalently, large intrinsic Q-factor $Q = \lambda_c / \gamma$ of the resonance determined by Eq. (1), requires the extraordinary translation symmetry (uniformity) of the waveguide.

Indeed, assuming $Q \geq 10^5$ (achievable for the photonic crystal and coupled microresonator waveguides [10, 11, 14, 15], FBG waveguides [16], and WGM waveguides [17]) and characteristic $\lambda_c \sim 1$ μm, we arrive at the STPC bandwidth $\Delta\lambda_{STPC}$ compared to $\gamma \sim 10$ pm. The observation of STPC also requires that the variation of the cutoff wavelength $\delta\lambda_c$ along the waveguide length caused by the waveguide nonuniformity should be much smaller than the bandwidth $\Delta\lambda_{STPC}$. Then, for the characteristic waveguide width $d \sim 10 - 100$ μm, the waveguide nonuniformity should be $\delta d \ll d/Q \sim 0.1 - 1$ nm. The latter condition is currently unachievable for the photonic crystal and ring microresonator structures. Favorably, the condition $\delta\lambda_c \ll \Delta\lambda_{STPC}$ is feasible for FBG waveguides [16] and WGM fiber waveguides [5, 17], which transmission losses are usually smaller than the losses of planar and photonic waveguides.

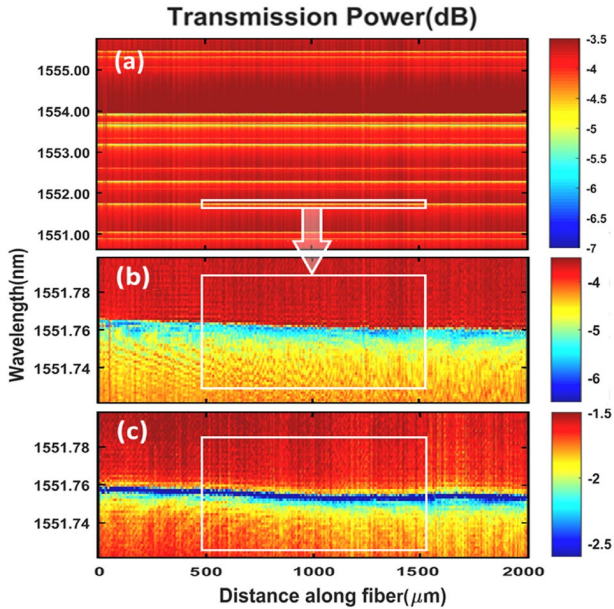

**Fig. 2.** (a) The spectrogram of the transmission power for the TF used in our experiment measured along its 2 mm length. (b) The magnified section of spectrogram (a) near the cutoff wavelength 1551.76 nm. (c) The spectrogram measured near the same cutoff wavelength as in (b) but with a smaller coupling strength between MF1 and TF.

The study presented in this Letter was motivated by our suspicion that, currently, a conventional optical fiber can serve as the best candidate for the accurate and versatile observation of the STPC. In fact, in addition to its exceptional uniformity and low loss, an optical fiber gives us a unique opportunity to investigate the propagation of WGMs which are localized near the fiber surface and therefore can be probed from the outside at any axial position along the fiber length. This allows us to investigate the STPC phenomenon in situ and for a tunable potential barrier.

Light propagating along the optical fiber is characterized by its azimuthal and radial quantum numbers, $m$ and $p$, and polarization [18]. For small $m$ and $p$, the proximity to the cutoff wavelengths is associated with losses caused by mode delocalization along the fiber cross-section [18] rather than slowing down and tunneling along the fiber axis. Alternatively, for WGMs with large $m$ and $p$ the delocalization and loss at the cutoff are negligible because the effective potential barrier along the radial direction responsible for these effects becomes large.

Our experimental setup is illustrated in Fig. 1(f). It consists of the test fiber (TF) positioned in direct contact and oriented transversely to the micrometer diameter waists (microfibers) MF1 and MF2 of the input-output biconical tapers 1 and 2. WGMs are evanescently launched in the TF by the MF1 and probed by MF2 separated from MF1 by distance $L$. Coupling between the input-output microfibers, MF1 and MF2, and the test fiber, TF, was adjusted by translation of the microfiber tapers along their axes, i.e., by changing the microfiber diameter at its contact point with the TF.

The theory of propagation of WGMs along a uniform TF coupled to the input-output microfibers illustrated in Fig. 1(f) was developed in Ref. [5]. In this theory, coupling of TF with MF1 and MF2 is determined by parameters $C_1$, $D_1$ and $C_2$, $D_2$, respectively. In the vicinity of a cutoff wavelength $\lambda_c$, the resonant transmission amplitudes along the MF1, $S_{11}(\lambda)$, and from the MF1 to MF2, $S_{12}(\lambda)$, are found as

$$S_{11}(\lambda) = S_{11}^{(0)} - iC_1 C_1^* G(\lambda, 0), \quad S_{12}(\lambda, L) = S_{12}^{(0)} - iC_1 C_2^* G(\lambda, L). \quad (2)$$

Here constants $S_{11}^{(0)}$ and $S_{12}^{(0)}$ determine the nonresonant part of these transmission amplitudes (ideally, $S_{11}^{(0)} = 1$ and $S_{12}^{(0)} = 0$) and

$$G(\lambda, L) = \frac{2i\beta(\lambda)\exp(i\beta(\lambda)L)}{(2i\beta(\lambda) + D_1)(2i\beta(\lambda) + D_2) - D_1 D_2 \exp(2i\beta(\lambda)L)}, \quad (3)$$

where $\beta(\lambda)$ is determined by Eq. (1) with $\beta_0 = 0$ and $K = 2^{3/2}\pi n_r \lambda_c^{-3/2}$ [5]. The experimentally measured group delay time is determined from Eq. (2) as [7]

$$\tau_{gr}(\lambda, L) = \frac{\lambda_c^2}{2\pi c}\text{Im}\left(\frac{\partial \ln(S_{12}(\lambda, L))}{\partial \lambda}\right), \quad (4)$$

where $c$ is the speed of light. For the case of $S_{12}^{(0)} = 0$ (i.e., in the absence of nonresonant coupling losses) and under the condition that coupling coefficients $D_1 \gg |\beta(\lambda)| \gg D_2$, we have $S_{12}(\lambda, L) = -C_1 C_2^* D_1^{-1} \exp(i\beta(\lambda)L)$. In this case, the reflection from the microfibers can be neglected and the WGM propagation is purely one-directional. Then the group delay time coincides with the classical delay $\tau_{cl}(\lambda, L) = 2\pi c \lambda_c^{-2} \text{Re}(\partial\beta(\lambda)/\partial\lambda^{-1})L$ $= n_r L \lambda_c^{1/2} [\delta + (\delta^2 + 1)^{1/2}]^{1/2} / 2c\gamma^{1/2}(\delta^2 + 1)^{1/2}$, where $\delta = (\lambda_c - \lambda)/\gamma$. In this case, the STPC region has the order of $\gamma$. In a more general situation, the effect of coupling between the input-output microfibers and TF cannot be neglected and the group delay time determined by Eq. (4) is affected by reflection of light at the contact points. Consequently, the STPC region can be much greater than $\gamma$.

The axial nonuniformity of an optical fiber (very small, though measurable [17]) and propagation loss of WGMs determine the practical limitations for the observation of the STPC. The effect of loss is simply described by parameters $\gamma$, $C_i$ and $D_i$ in Eqs. (1)-(4). The characteristic length $L_0$ of the fiber segment of our interest is determined by the attenuation of WGM power caused by this loss and is estimated from $\exp(-2|\beta(\lambda_c)|L_0) \sim 1$. For a silica fiber with refractive index $n_r = 1.45$ at the cutoff wavelength $\lambda_c = 1.5$ μm, the measured WGM Q-factor is $Q \sim 10^7$ (see below), so that $\gamma = \lambda_c / Q = 0.15$ pm and therefore $L_0 = \lambda_c^{3/2} \gamma^{-1/2} (4\pi n_r)^{-1} \sim 300$ µm. Next, the effect of fiber nonuniformity can be estimated as follows. The length of a fiber segment of our concern is a few $L_0$, i.e., several hundreds of microns. The effective radius variation (ERV) of the optical fiber along this length is usually linear, $\Delta r(z) = \alpha z$ with slope $\alpha$ [17]. A WGM excited by a microfiber (Fig. 1(f)) is reflected at the turning point [5] if this point is separated from the microfiber by the distance of the order of $L_0$. Otherwise, it will be attenuated so that the reflected part of this mode can be neglected. Variation of the cutoff wavelength along the fiber is expressed through ERV $\delta r(z)$ as $\delta \lambda_c(z) = \delta r(z) \lambda_c / r_0$ where $r_0$ is the fiber radius [18]. To avoid the effect of reflection, we have to require that the separation of the WGM wavelength from the cutoff wavelength $\Delta \lambda = |\lambda - \lambda_c|$ is greater than $\alpha \lambda_c L_0 / r_0$. In the experiment described below, we have $\alpha = 2 \cdot 10^{-7}$ which yields $\Delta \lambda > 1$ pm.

In our experiment, the TF was a segment of a standard silica optical fiber with radius $r_0 = 62.5$ µm which was mechanically stripped from coating and rinsed in isopropanol. The uniformity of TF was characterized by scanning the MF1 along its 2 mm length [17] in the absence of MF2. The spectrogram of the transmission power shown in Fig. 2(a) (a linear combination of transmission powers $P_{11}(\lambda, z) = |S_{11}(\lambda)|^2$ for TE and TM polarizations) was measured with 4 µm resolution along the fiber axis $z$ and 1.3 pm resolution along wavelength $\lambda$ using Luna Optical Vector Analyzer. In this plot, the resonant transmission power dips correspond to the WGM cutoff wavelengths. To ensure the visibility of larger number of cutoff wavelengths, which correspond to different quantum numbers, $m$ and $p$, and two polarizations, the coupling between MF1 and TF was maximized. Consequently, the off-resonant transmission power was decreased due to the increased coupling losses. For the ideally uniform fiber, the transmission power $P_{11}(\lambda, z)$ does not depend on $z$ and the positions of dips are exactly horizontal. Very small variation of the cutoff wavelengths along the TF length is seen in Fig. 2(b), which magnifies Fig. 2(a) along the vertical axis near the cutoff wavelength $\lambda_c = 1551.76$ nm. In order to reduce the apparatus noise of our measurements and verify repeatability, we characterized the same TF segment with a smaller coupling between MF1 and TF, which corresponded to a larger diameter of MF1 at contact points. The measured spectrogram shown in Fig. 2(c) demonstrates smaller noise and more accurate ERV characterization than that in Fig. 2(b). Notice small shift in the position of $\lambda_c$ in Fig. 2(b) and (c) due to slightly different environmental temperatures. The ERV variation $\delta r$ can be determined from the cutoff wavelength variation $\delta \lambda_c$ by the rescaling relation $\delta r = \delta \lambda_c r_0 / \lambda_c$. In addition to the original nonuniformity of the optical fiber, the detected ERV is caused by the thermal fluctuations during the 2-hour long characterization of the TF and the fiber surface contamination (e.g., by residuals of the coating material [19]). From Fig. 2(c), the cutoff wavelength variation along the 1 mm of the TF is $\delta \lambda_c \sim 5$ pm which corresponds to the ERV $\delta r \sim 0.2$ nm, while the slope of ERV is estimated as $\alpha \sim 0.2$nm/1mm$=2 \cdot 10^{-7}$. For comparison, $\delta \lambda_c \sim 5$ pm corresponds to the environmental temperature variation $\delta T \sim n_r \delta \lambda_c / \zeta \lambda_c \sim 0.5$ K (here $\zeta = 1.2 \cdot 10^{-5}$ K$^{-1}$ is the thermo-optic coefficient of silica). Therefore, the detected ERV can be partly due to the effect of temperature variation. However, comparison of spectrograms shown in Figs. 2(b) and 2(c), which provide remarkably close characteristic ERV suggest that these figures present a good estimate for the actual ERV of our TF.

Parameters $C_1$, $C_2$, $D_1$, $D_2$, $S_{11}^{(0)}$ and $S_{12}^{(0)}$ which determine the coupling between the TF and microfibers MF1 and MF2, and the resonance width $\gamma$, which characterize the intrinsic losses of the TF, were determined by fitting the resonance spectrum of the TF measured with MF1 (MF2) while MF2 (MF1) was disconnected. In our experiment, we tuned MF1 and MF2 to have the same coupling parameters in the vicinity of cutoff wavelength $\lambda_c = 1555.76$ nm. The fitting method similar to the one described previously [5] yields $C_1 = C_2 = 0.11 \ \mu m^{-1/2}$, $\text{Im} D_1 = \text{Im} D_2 = 0.008 \pm 0.002 \ \mu m^{-1}$, $\text{Re} D_1, \text{Re} D_2 < 0.001 \mu m^{-1}$, and $\gamma < 0.1$ pm. We set $S_{12}^{(0)} = 0$, which follows from the exponentially small values of $S_{12}(\lambda, L)$ in the tunneling region of spectra (i.e., for $\lambda > \lambda_c$ within the STPC) and for relatively large separation $L$ between MF1 and MF2 (see Fig. 3(a) below).

Next, the slow propagation and tunneling of WGMs along the TF was characterized with both MF1 and MF2 (Fig. 1(f)). Specifically, the input MF1 was fixed at the position of $z = 1000 \ \mu m$ and the output MF2 was translated from $z = 400 \ \mu m$ to $z = 1600 \ \mu m$, i.e., within the length of the white rectangle in Figs. 2(b) and (c), with the measurement steps of 4 µm. To avoid the contact between MF1 and MF2 during the translation of the MF2 along the TF, both microfibers were U-shaped as illustrated in Fig. 1(f). In our experiment, we measured and theoretically analyzed the spectrograms of transmission power and group delay of light in the vicinity of cutoff wavelength $\lambda_c = 1555.76$ nm outlined by white rectangles in Fig. 2. The coupling parameters indicated were equal to the coupling parameters of the spectrogram in Fig. 2(b). As noted above, Fig. 2 shows the spectrograms of the full transmission power through the MF1, which includes the cutoff wavelengths of both polarizations. Separation of these polarizations (not performed in Fig. 2) is possible by diagonalization of the measured Jones matrix [20]. In contrast, the measurement of *resonant* transmission amplitude of light with a single polarization, $S_{12}(\lambda)$, does not require the diagonalization of the Jones matrix since the contribution of non-resonant light transmitted from MF1 to MF2 is negligible. In spite of this simplification, the measurement of the transmission power $P_{12}(\lambda, z) = |S_{12}(\lambda)|^2$ and group delay time $\tau_{gr}(\lambda, L)$ defined by Eq. (4) was challenging because of the small values of $P_{12}(\lambda, z)$ compared with the noise background of our spectrum analyzer (this problem can be directly resolved in the future with more accurate measurements). Notice that, as follows from the uniformity of spectrograms in Fig. 2, the noise caused by shaking of the microfiber in the process of measurement is negligible. To reduce noise, the spectrograms of recorded transmission power and group delay shown in Fig. 3(a) and (b) were averaged over ten successive measurements at each location as well as over 7 measurement points (28 µm) along the fiber axis and two points (2.6 pm) in spectrum. The data shown in Fig. 3(a) and (b) was collected during 5 hours of continuous measurements.

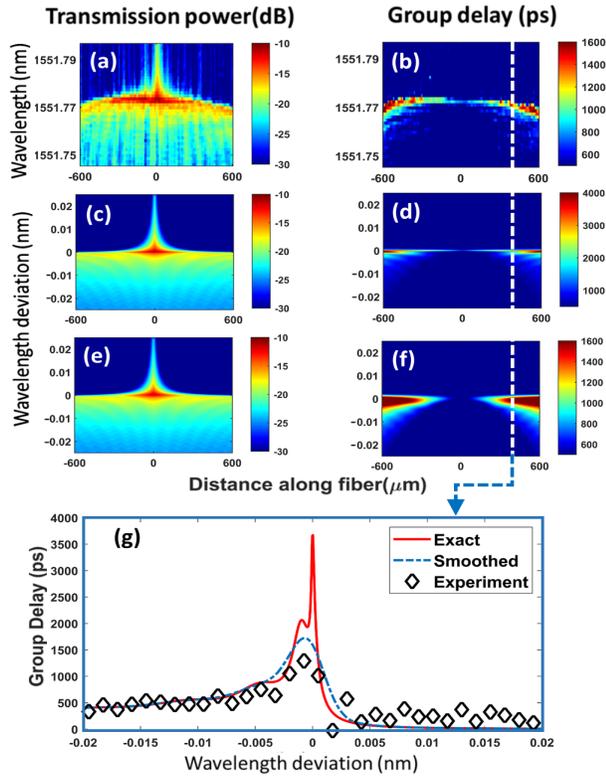

Fig. 3. Spectrograms of transmission power $|S_{12}(\lambda,L)|^2$ and group delay $\tau_{gr}(\lambda,L)$ in the vicinity of the cutoff wavelength $\lambda_c = 1555.76$ nm. The axial coordinate $z=0$ is where the MF1 is located, while MF2 is translated from $z=-600$ μm to $z=600$ μm. (a) and (b) – experimental; (c) and (d) – theoretical calculated from Eqs. (1)-(4); (e) and (f) – theoretical averaged over 28 μm along the fiber axis and over 2.6 pm in spectrum. (g) compares the experimental group delay spectrum at $L=400$ μm (rhombs) with exact (solid red line) and averaged (dash blue line) theoretical spectra.

For comparison, Fig. 3(c) and (d) show the spectrograms of the transmission power and group delay calculated from Eqs. (1)-(4) using the measured coupling parameters. In turn, Fig. 3(d) and (f) are the spectrograms of Fig. 3(c) and (d) averaged over the same axial length and spectral width as in our experiment. Comparison of experimental data in Fig. 3(a) and (b) with numerical simulations shown in Fig. 3(e) and (f) demonstrates good agreement. In particular, Fig. 3(g) compares the experimental spectrum with exact and smoothed theoretical spectra for the separation between MF1 and MF2 equal to 400 μm. The transmission power spectrograms in Fig. 3 clearly distinguish the tunneling region above the cutoff wavelength and slow light propagation region below it. The complex behavior of the group delay in the theoretical spectrograms (which is smoothed out after averaging) and oscillations of the transmission power in the slow light region is explained by the reflection of WGMs from the microfibers. A small bending down of the spectrograms with increasing the separation between MF1 and MF2 in the experimental spectrograms is presumably caused by the temperature variation during the 5-hour time of data collection.

In summary, we have demonstrated that a conventional optical fiber, being exceptionally uniform and low loss, can serve as a unique means for the investigation of the crossover between slow and tunneling light. In contrast to the devices used previously for demonstration of the STPC phenomenon (e.g., photonic crystals, fiber Bragg gratings, and coupled microresonators), an optical fiber allows us to investigate it in situ and for tunable potential barriers. We suggest that, with a more accurate detection of the transmission spectra currently feasible, different features of the STPC phenomenon can be experimentally demonstrated. In particular, by varying the coupling parameters between the test fiber and input-output microfibers, it is possible to exclude the effect of reflections at the contact points and arrive at the pure tunneling and slow light propagation. Our setup illustrated in Fig. 1(f) can be directly tuned to realize the latter condition as well as other coupling conditions under interest.

**Funding.** Engineering and Physical Sciences Research Council (EPSRC) (EP/P006183/1); Horizon 2020 MCSA COFUND MULTIPLY (H2020 GA 713694);

**Disclosures.** The authors declare no conflicts of interest.